\documentclass[fleqn,twoside]{article}
\usepackage[headings]{espcrc2}

\readRCS
$Id: espcrc2.tex,v 1.2 2004/02/24 11:22:11 spepping Exp $
\usepackage{graphicx}
\usepackage[figuresright]{rotating}

\newcommand{\ens}{\epsilon_{NS}}
\newcommand{\es}{\epsilon_{S}}

\newcommand{\Veff}{{\cal V}}    
\newcommand{\Aeff}{{\cal A}}

\newcommand{\AmS}{{\protect\the\textfont2
  A\kern-.1667em\lower.5ex\hbox{M}\kern-.125emS}}

\title{Can we probe right-handed charged quark currents?}
\author{Micaela Oertel \address{LUTH, Observatoire de Paris-Meudon,5 place Jules Janssen, 92195 Meudon, France}%
\thanks{Talk based on work in collaboration with V. Bernard, E. Passemar, and J.Stern.}
}       

\begin{document}

\begin{abstract}
Different scenarios of new physics beyond the standard model give rise
to a direct coupling of right-handed quarks to $W$ bosons. We will
discuss how the prediction of the Callan-Treiman low energy theorem
for the scalar $K\pi$ form factor in combination with recent data on
$K^L_{\mu 3}$ decays can serve as a stringent test of the standard
model (SM) and
possible extensions, in particular right-handed charged quark
currents (RHCs). In addition, we will comment on the impact of 
hadronic tau decay data on RHCs.
\vspace{1pc}
\end{abstract}

\maketitle
\section{Introduction}
Within the (not quite) decoupling low energy effective theory (LEET)
scenario discussed by Jan Stern~\cite{talkjan,HS1}, the only two operators
appearing at NLO modify the charged current (CC)
interaction. We will discuss whether it is possible to experimentally
constrain the corresponding couplings. We will focus on the light
quark sector where we will have to cope with the problem that it is
not easy to disentangle the effects of non-standard electroweak
couplings from (non-perturbative) QCD effects. We will see that in
this respect $K^L_{\mu 3}$ decays present a stringent test of the SM
and possible extensions giving rise to right-handed charged quark
currents (RHCs).

In this talk we will consider a universal modification of the
quark CC interaction leading to the following effective
couplings at NLO for the vector and axial quark current, respectively:
\begin{eqnarray}
\Veff_{\mathit{eff}}^{ij}&=&(1+\delta) V_L^{ij}+\epsilon
V_R^{ij}+\mathrm{NNLO}\nonumber \\
\Aeff_{\mathit{eff}}^{ij}&=&-(1+\delta) V_L^{ij}+\epsilon V_R^{ij}+\mathrm{NNLO}~,
\label{effective couplings}
\end{eqnarray}
where $V_L$ and $V_R$ are two a priori independent quark mixing
matrices and the two parameters $\delta$ and $\epsilon$ measure the
departure from the SM. The leptonic CC will not be
modified~\cite{HS1}. Within the LEET scenario, the two parameters
$\delta$ and $\epsilon$ can be related to the two NLO operators, but
the same structure is shared by many models, as for example left-right
symmetric extensions of the SM.

In tree level processes, we will encounter in the light quark sector, apart 
from $\delta$, the following two combinations:
\begin{equation}
\ens= \epsilon\ \mathrm{Re}
\Bigl{(}\frac{V_R^{ud}}{V_L^{ud}}\Bigr{)}, \quad
\es =\epsilon\ 
\mathrm{Re} \Bigl{(}\frac{V_R^{us}}{V_L^{us}}\Bigr{)}~.
\label{epsilon}
\end{equation}
measuring the amount of $\bar u d$ and $\bar u s$ RHCs, respectively. 
We will in turn discuss the information we obtain on these 
parameters from $K^L_{\mu 3}$ and hadronic tau decays. 
\section{$K^L_{\mu 3}$ decays}
The hadronic matrix element describing
the $K^{0}_{\mu 3}$ decay can be written in terms of two form factors:
\begin{eqnarray}
&& \langle \pi^-(p') | \bar{s}\gamma_{\mu}u | K^0(p)\rangle =  \nonumber \\ 
&& (p'+p)_\mu\  f^{K^0\pi^-}_+ (t) + (p-p')_\mu\  f_-^{K^0\pi^-} (t)~,\qquad         
\label{hadronic element}
\end{eqnarray}
where $t=(p'-p)^2$. 
We will concentrate
on the normalized scalar form factor    
\begin{equation}
f(t)=\frac{f^{K^0\pi^-}_+ (t)}{f^{K^0\pi^-}_+(0)} + \frac{t}{m^2_{K^{0}} - m^2_{\pi^{-}}} \frac{f^{K^0\pi^-}_-(t)}{f^{K^0\pi^-}_+(0)}
\label{defnffactor}
\end{equation}
The Callan-Treiman low-energy theorem (CT)~\cite{Dashen:1969bh} fixes
the value of $f(t)$ at the point
$t=\Delta_{K\pi}=m_{K^{0}}^2-m_{\pi^-}^2$ in the $\mathrm{SU}(2)\times
\mathrm{SU}(2)$ chiral limit. We can write
\begin{equation}
C=f(\Delta_{K\pi})=\frac{F_{K^+}}{F_{\pi^+}}\frac{1}{f_{+}^{K^0\pi^-}(0)}+  
\Delta_{CT},
\label{C}
\end{equation}
where the CT discrepancy $\Delta_{CT}$ defined by Eq.~($\ref{C}$) is
expected to be small and eventually calculable in $\chi PT$.  It is
proportional to $m_u$ and/or $m_d$. In the limit $m_d=m_u$ at the NLO
in $\chi PT$ one has for the CT discrepancy
$\Delta_{CT}^{\mathrm{NLO}}= - 3.5 \times
10^{-3}$~\cite{Gasser:1984ux}. We will focus the discussion on the
neutral kaon mode since the analysis of the charged mode is subject to
larger uncertainties related, in particular, to $\pi^0\eta$
mixing~\cite{Bernard:2006gy}.

At low energies the form factor can be parameterized accurately in
terms of only one parameter, $\ln C$, in a model independent way. To
that end we employ a twice subtracted dispersion relation. One usually
assumes that $f(t)$ has no zeros. In that case we can
write~\cite{Bernard:2006gy}:
\begin{equation}
f(t)=\exp\Bigl{[}\frac{t}{\Delta_{K\pi}}(\ln C- G(t))\Bigr{]}~,
\label{Dispf}
\end{equation}
where 
\begin{eqnarray*}
G(t)& =& 
\frac{\Delta_{K\pi}(\Delta_{K\pi}-t)}{\pi} \\ && \times \int_{t_{\pi K}}^{\infty}
\frac{dx}{x}\frac{\phi(x)} {(x-\Delta_{K\pi})(x-t-i\epsilon)}~.
\end{eqnarray*}
$t_{\pi K}$ is the threshold of $\pi K$ scattering and $\phi(t)$ is
the phase of $f(t)$. At sufficiently low energies this phase should
agree, due to Watson's theorem, with the $s$-wave, $I = 1/2, K\pi$
scattering phase, $\delta_{K\pi}(t)$. We take $\phi(t) =
\delta_{K\pi}(t) $ up to an energy of $1.67~\mathrm{GeV}$. In this
domain, as observed experimentally~\cite{Estabrooks:1977xe}, the
scattering amplitude is to a very good approximation elastic, and
$\delta_{K\pi}$ is known precisely from a Roy-Steiner
analysis~\cite{Buettiker:2003pp}.  Following
Brodsky-Lepage~\cite{Lepage:1979zb}, asymptotically the phase should
reach $\pi$. In between the elastic and the asymptotic region we will
assume $\phi(t) = \pi \pm \pi$.  In principle it is possible to infer
the phase of the form factor above the elastic region using a somewhat
model dependent Omnes-Mushkelishvili construction as presented in
Ref.~\cite{Jamin:2001zq}. In any case, due to the two subtractions the
dispersive integral converges rapidly, keeping the error arising from
the uncertainties in the high-energy behavior of the phase small. We
nevertheless checked that in the low-energy region the phase of
Ref.~\cite{Jamin:2001zq} reproduces our function $G(t)$ within errors.
It can be observed
that within the whole physical region the value of $G(t)$ stays at
least a factor of five smaller than the expected value of $\ln
C$~\cite{Bernard:2006gy}. We include uncertainties on the extraction
of $\delta_{K \pi}$ and the high-energy behavior into the error on $G(t)$.

Our dispersive representation allows to obtain the slope, $\lambda_0$,
and the curvature, $\lambda'$, of the form factor at $t = 0$ in terms
of only one parameter, $\ln C$:
\begin{eqnarray}
\lambda_0 &=& \frac{m_{\pi}^2}{\Delta_{K \pi}}(\ln C - G(0)
) ~,\label{slope}\\
\lambda' &=& \lambda^2  - 2 \frac{m_{\pi}^4}{\Delta_{K\pi}}
G'(0)~,
\label{curvature}
\end{eqnarray}
where the Taylor expansion of the form factor has been written as follows:
$f(t) = 1 + \lambda_0 \frac{t}{m_{\pi}^2} + \frac{1}{2} \lambda'
(\frac{t}{m_{\pi}^2})^2  + \ldots~.$
What is the value of $\ln C$? We can express $C$ in terms of the
measured branching ratio
Br~$K^+_{l2}(\gamma)/\pi^+_{l2}(\gamma)$~\cite{Jamin:2006tj},
the inclusive decay rate $K^L_{e3} (\gamma)$
~\cite{fplus}, and the value
of $|\Veff_{\mathit{eff}}^{ud}|$ known from superallowed $0^+\to 0^+$ nuclear
$\beta$-decays~\cite{Marciano:2005ec} as
$C = B_{\mathit{exp}} \, r + \Delta_{CT}\label{lnc}~,$
with \[ B_{\mathit{exp}} = \Bigl{|}\frac{F_{K^+} \Aeff_{\mathit{eff}}^{us}}
{F_{\pi^+} \Aeff_{\mathit{eff}}^{ud}}\Bigr{|}
\frac{1}{|f_+^{K^0\pi^-}(0)
\Veff_{\mathit{eff}}^{us}|}|\Veff_{\mathit{eff}}^{ud}|\] and \[r = \Bigl{|}\frac{\Aeff_{\mathit{eff}}^{ud} \Veff_{\mathit{eff}}^{us}}{\Veff_{\mathit{eff}}^{ud}
\Aeff_{\mathit{eff}}^{us}}\Bigr{|}~.\] This gives to first order in $\epsilon$
\begin{eqnarray}
\ln C  &=& 0.2183 \pm 0.0031 + \tilde\Delta_{CT} + 2 (\epsilon_S -
\epsilon_{NS})\nonumber\\
 &=& 0.2183 \pm 0.0031 + \Delta\epsilon
\label{lncexp}
\end{eqnarray}
where $\tilde \Delta_{CT} = \Delta_{CT}/B_{\mathit{exp}}$.
$\es$ and $\ens$ have been defined in Eq.~(\ref{epsilon}).

Assuming SM weak interactions, i.e., $\es = \ens = 0$, we
obtain the
following very precise prediction for $\lambda_0$
(cf. Eqs.~(\ref{slope}),(\ref{lncexp})):
\begin{equation}
\lambda_0 = 0.01524 \pm 0.00044 + 0.0686 \Delta_{CT}~.
\label{lambda0}
\end{equation}
This value can be compared with the slope parameter
$\lambda_{\mathit{exp}}$ measured in
$K^L_{\mu 3}$ decay experiments. It has to be stressed that the
measured slope parameter cannot be directly interpreted as the slope
of the form factor at $t = 0$ since it is determined from a global fit
to the measured decay distributions employing a linear
parametrization, $f(t) = 1 + \lambda_{\mathit{exp}} t/m_\pi^2$,
for the form factor.  
To better illustrate the problem, let us define an effective slope by
$
f(t) = 1 + \lambda_{\mathit{eff}}(t) \ t/m_\pi^2~.
$
Since the curvature of the form factor is positive,
$\lambda_{\mathit{eff}}(t)$ is a monotonically rising function of $t$.
This means that $\lambda_{\mathit{exp}} \ge \lambda_0$, i.e., the
measured slope parameter represents an upper bound for the value of
$\lambda_0$. 

Comparing the published value of the KTeV collaboration,
$\lambda_{\mathit{exp}} = 0.01372 \pm 
0.00131$~\cite{Alexopoulos:2004sy}, and the preliminary value of NA48,
$\lambda_{\mathit{exp}} = 0.0120 \pm 0.0017   $~\cite{NA48}, with the
SM prediction in Eq.~(\ref{lambda0}), this indicates that we need a
correction on the percent level, whereas estimates of the
Callan-Treiman correction within $\chi PT$ give $\tilde\Delta_{CT}
\sim 10^{-3}$~\cite{Gasser:1984ux}. 

Thus, the Callan-Treiman low-energy theorem, in combination with
measurements of the scalar $K\pi$ form factor offers a very
interesting test of the SM and possible new physics. 

If we admit that the charged current interaction gets modified
and that we have RHCs, we get an additional correction sensitive to
$\es - \ens$, cf. Eq.~(\ref{lncexp}). $\es$ and $\ens$ 
represent the strengths of $\bar ud$ and $\bar
us$ RHCs, respectively. 
It should be mentioned on the one hand that
RHCs can escape detection in $K^L_{\mu 3}$ decays, if the right-handed
mixing matrix is aligned with the left-handed CKM matrix, i.e., $\es =
\ens$. On the other hand, it is possible that $\es$ gets enhanced by
an inverse hierarchy of mixing matrix elements in the right-handed
sector compared with the left-handed mixing matrix~\cite{talkjan}. If
we take the specific framework of the LEET~\cite{talkjan}, we expect
from power counting arguments $\epsilon$ to be of the order of
percent. Therefore $\ens$ cannot be much larger than 0.01, whereas
$\es$ could be larger. 

\begin{figure}[t!]
\vspace{0.1cm}
\begin{center}
\includegraphics*[scale=0.5]{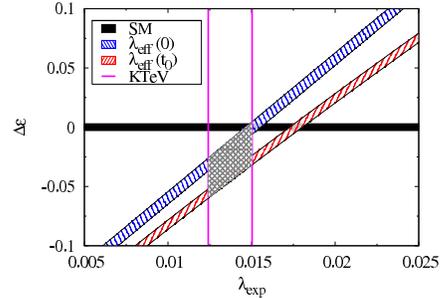}
\caption{\it Impact of KTeV data on RHCs. Horizontal line: SM, $\Delta \epsilon = \tilde{\Delta}_{CT} = \pm~0.0028$,
vertical lines: KTeV measurements of $\lambda_0$. Top hatched curve: $\lambda_{\mathit{exp}}=\lambda_{\mathit{eff}}(0)$  and bottom
hatched curve: $\lambda_{\mathit{exp}}=\lambda_{\mathit{eff}}(t_0)$ with uncertainties from branching ratios and from G(t) added in quadrature.}    
\label{figlambda}
\end{center}
\end{figure}

Let us now infer the value of $\Delta\epsilon$ from the data on
$\lambda_{\mathit{exp}}$. Unfortunately the determination of $\ln C$
from $\lambda_{\mathit{exp}}$ is subject to large parametrization
uncertainties. As explained above, the interpretation of
$\lambda_{\mathit{exp}}$ in terms of the slope of the form factor is not
clear and the measured slope parameter represents only an upper bound for
$\lambda_0$. 
We therefore only get an upper bound for $\Delta\epsilon$: 
\begin{eqnarray*}
\Delta\epsilon_{\mathit{max}} &=& -0.0178 \pm 0.0156 \pm
0.0040 ~[\mathrm{KTeV}]~,\\
\Delta\epsilon_{\mathit{max}} &=& -0.0379 \pm 0.0201\pm
0.0040~ [\mathrm{NA48}]~.
\end{eqnarray*}
The first error corresponds to the experimental error on
$\lambda_{\mathit{exp}}$ and the branching ratios. The second one
indicates the error on $G(t)$ inherent to our dispersive
representation. The situation is illustrated for the KTeV data in
Fig.~\ref{figlambda} where we display the value of $\Delta\epsilon$
in terms of $\lambda_{\mathit{exp}}$. The upper hatched curve
indicates the result choosing $\lambda_{\mathit{exp}} =
\lambda_{\mathit{eff}}(0)$, the lower hatched curve
$\lambda_{\mathit{exp}} = \lambda_{\mathit{eff}}(t_0)$.  Errors on the
branching ratios and on $G(t)$ have been added in quadrature. The
vertical lines indicate the KTeV measurement, whereas the
horizontal line corresponds to the SM case with $\Delta\epsilon =
\tilde\Delta_{CT} = \pm 0.0028$~\cite{Gasser:1984ux}. Including the
parametrization uncertainty as an error, we obtain $\Delta\epsilon =
-0.03 \pm 0.03$ (KTeV data) as indicated by the gray shaded area. The
same procedure leads to $\Delta\epsilon = -0.05 \pm 0.03$ using the
preliminary NA48 data. These values are perfectly consistent with
the expectation from the LEET, but pointing to an enhancement of
$\es$.

From the above discussion it is clear that the uncertainty due to the
difficulties in interpreting the measured slope parameter dominates.
The dispersive representation we proposed for the form
factor~\cite{Bernard:2006gy}, allows for an accurate description
within the whole physical region in terms of only one parameter, $\ln
C$. As explained above, a direct measurement of $\ln C$ can prove 
very important in testing the SM and possible physics beyond the SM.
We discussed in which way a possible discrepancy between the SM, as
indicated by Eq.~(\ref{lambda0}), and the measured slope parameter,
could be explained as an effect of RHCs.

\section{Hadronic tau decays}
The hadronic tau decays are semileptonic decays involving the charged
current.  Even though the different analyses of these decays done so
far \cite{Davier:2005xq} have not yet reported any evidence of physics
beyond the SM it seems interesting to reconsider them in
the light of our generalization of the electroweak charged current. 

For our analysis
we have considered the normalized total hadronic width given by the ratio 
\begin{equation}
R_{\tau,i} = \frac{\Gamma(\tau^-\to \nu_\tau \mathrm{hadrons}
(\gamma))}{\Gamma(\tau^- \to \nu_\tau e^-\bar\nu_e)}~,
\label{rtau}
\end{equation}
where $i$ can be $V,A$ or $S$ signifying that we are looking at the
vector, axial or strange channel, respectively. Additional information
is provided by spectral moments which explore
the invariant mass distribution of final state hadrons~\cite{Davier:2005xq}.

The theoretical description of these ratios can be separated into
several parts: the electroweak part, a perturbative QCD part and
non-perturbative contributions, eventually calculable within the
operator product expansion~\cite{Braaten:1991qm,LeDiberder:1992fr}.
The QCD corrections are functions of several QCD parameters:
$\alpha_s$, quark masses and
non-perturbative condensates. 

Details of the analysis will be presented elsewhere~\cite{BOPS06}. Here we
only want to summarize the main results.  An important point is that
present data only allow for putting constraints on the parameter
$\ens$ --in particular from the non-strange, $V+A$ channel-- and the
combination $\delta + \ens$, showing up in the strange channel. In
contrast to $\es$, which can be enhanced if there is no strong Cabbibo
suppression in the right-handed sector, we do not expect any
enhancement for these two quantities. This means that we are looking
for effects on the 1\% level or even below.  In fact, our knowledge of
the QCD parameters entering the analysis has not yet reached a
sufficient precision to determine effects of this order. Our
conclusion is that presently the analysis of tau decay data does not
exclude the presence of RHCs, in particular for the moment there is no
evidence for an inconsistency with the expansion within the LEET, but
on the other hand it does not allow for a quantitative determination
of $\ens$ and $\delta+\ens$.

\end{document}